\PassOptionsToPackage{table}{xcolor}
\documentclass[manuscript,screen]{acmart}
\AtBeginDocument{%
  }

\usepackage{algorithmic}
\usepackage{algorithm}

\usepackage[table]{xcolor} 
\usepackage{multirow}
\usepackage{multicol}
\usepackage{array}
\usepackage{makecell}
\setcellgapes{0pt}    

\setcopyright{acmlicensed}
\copyrightyear{2018}
\acmYear{2018}
\acmDOI{XXXXXXX.XXXXXXX}
\acmConference[Conference acronym 'XX]{Make sure to enter the correct
  conference title from your rights confirmation email}{June 03--05,
  2018}{Woodstock, NY}
\acmISBN{978-1-4503-XXXX-X/2018/06}

\begin{document}

\title{SynMOE: A Syntax-Aware MOE-Guided Approach for PEFT-Based Code Intelligence Learning}


\author{Shuo Liu}
\authornote{Shuo Liu and Hao Ning contributed equally to this work and are co-first authors.}
\affiliation{%
  \institution{City University of Hong Kong}
  \city{Hong Kong}
  \country{China}}
\email{sliu273-c@my.cityu.edu.hk}

\author{Hao Ning}
\authornotemark[1]
\affiliation{%
  \institution{Shandong University}
  \city{Qingdao}
  \country{China}}
\email{hning@mail.sdu.edu.cn}

\author{Jacky Keung}
\affiliation{%
  \institution{City University of Hong Kong}
  \city{Hong Kong}
  \country{China}}
\email{Jacky.Keung@cityu.edu.hk}

\author{Zhen Yang}
\authornote{Zhen Yang is the sole corresponding author.}
\affiliation{%
  \institution{Shandong University}
  \city{Qingdao}
  \country{China}}
\email{zhenyang@sdu.edu.cn}

\author{Xiao Yu}
\affiliation{%
  \institution{Zhejiang University}
  \city{Hangzhou}
  \country{China}}
\email{xiao.yu@zju.edu.cn}

\author{Xing Hu}
\affiliation{%
  \institution{Zhejiang University}
  \city{Hangzhou}
  \country{China}}
\email{xinghu@zju.edu.cn}

\author{Yicheng Sun}
\affiliation{%
  \institution{City University of Hong Kong}
  \city{Hong Kong}
  \country{China}}
\email{yicsun2-c@my.cityu.edu.hk}

\renewcommand{\shortauthors}{Liu et al.}

\begin{abstract}

Parameter-Efficient Fine-Tuning (PEFT) has emerged as an important approach for adapting Language Models (LMs) in downstream tasks. However, in the domain of code intelligence, the performance of most PEFT methods relatively drops compared to Full fine-Tuning (FT), hindering their broader applicability. One of the main reasons is that current PEFT methods are generally based on assumptions about the low-rank nature or sparsity of parameter updates, lacking explicit modeling of code-related information. To tackle this limitation, we propose SynMOE, a syntax-aware MOE-guided approach for PEFT on code intelligence learning. Specifically, SynMOE contains two stages: syntactic information extraction and syntax-aware MOE-guided adaptation. It first obtains the syntactic labels of source code from Abstract Syntax Trees (ASTs), and then aligns the labels with the input tokens of code LMs. In the second stage, syntactic labels are integrated into the MOE module, participating in the routing decision of MOE for a more powerful syntactic-aware token representation. We evaluate the effectiveness of the proposed approach by applying it to two typical PEFT methods, LoRA and $(IA)^3$, across three diverse code intelligence tasks. Experimental results show that SynMOE achieves consistent and significant improvements. Even compared to FT, SynMOE achieves comparable or superior performance while being more computationally efficient. In addition, we also conduct ablation studies, hyperparameter analysis, and evaluation in the low-resource scenario to provide a comprehensive investigation of the proposed approach.

\end{abstract}




\maketitle

\section{Introduction}
Code Language Models (Code LMs) have demonstrated significant effectiveness across a wide range of code intelligence tasks, such as code summarization \cite{yang2021multi,yang2026r, yang2023significance}, code clone detection \cite{tao2022c4,liu2025can}, and code completion \cite{guo2024ft2ra, guo2023longcoder}. One of the mainstream approaches to adapting code LMs to downstream tasks is Full fine-Tuning (FT), but it suffers from high computational costs. As the scale of code LMs grows from millions to billions of parameters, this approach becomes impractical for many practitioners. Recently, Parameter-Efficient Fine-Tuning (PEFT) has emerged as a promising alternative to FT. It updates only a small subset of parameters while keeping the others frozen, making it particularly suitable for scenarios with limited computational resources and thus attracting much attention. 

\begin{figure}[t]
	\centering
        \setlength{\abovecaptionskip}{0.cm}
		\includegraphics[width=0.8\textwidth]{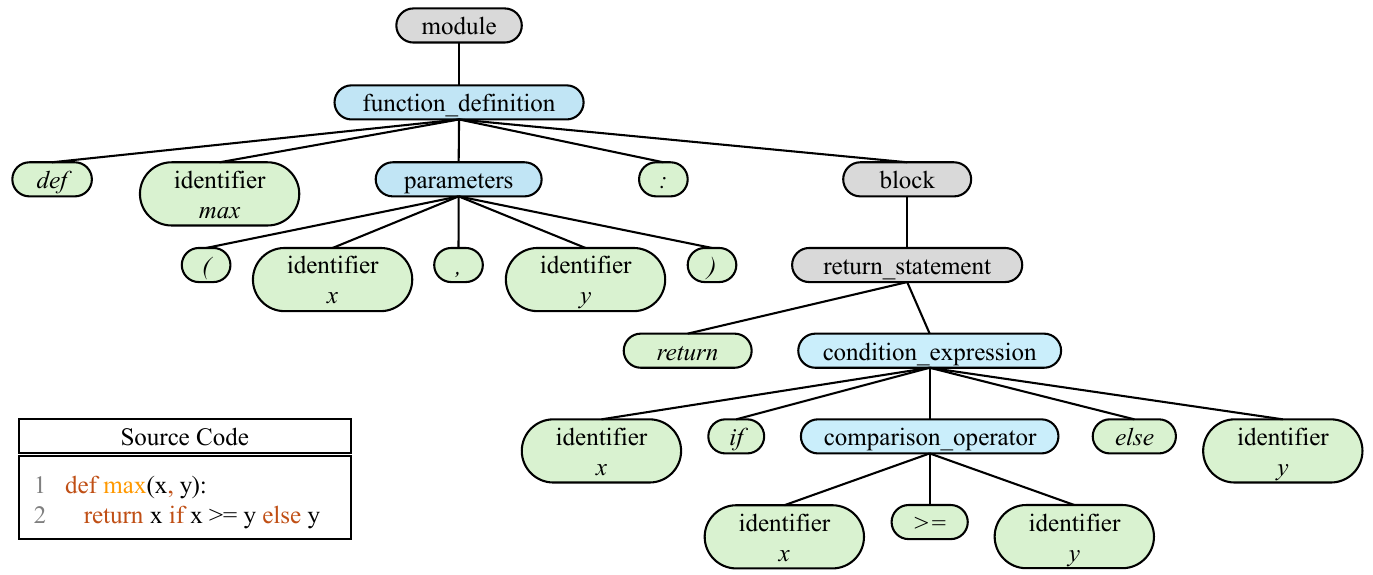}
	  \caption{A Python code snippet and its corresponding AST.}\label{ast-figure}
\vspace{-0.4cm}
\end{figure}

Despite these superiorities, previous studies \cite{liu2023empirical, liu2024delving, liu2025empirical} show that PEFT methods generally exhibit relatively lower performance than FT in the context of code intelligence tasks, such as code translation and commit message generation. We argue that vanilla PEFT methods are generally based on assumptions about the low-rank nature or sparsity of parameter updates, yet they lack explicit modeling of code-related information, making their application to code domains inherently suboptimal. Figure \ref{ast-figure} demonstrates a Python code snippet and its Abstract Syntax Tree (AST), which aims to return the maximum value between \texttt{x} and \texttt{y}. Human developers grasp this code effortlessly, largely because they intuitively assign different semantic roles to keywords, operators, and variables rather than viewing them as a flat sequence of text. For instance, \texttt{x} acts as a parameter in the parameter list (Line 1), while serving as a return value under the condition expression and as a comparison operand under the comparison expression (Line 2), which are mainly the observations syntactically elicited from the associated AST. As such, integrating this syntactic information can intuitively enhance the code comprehension capabilities of code LMs, particularly in PEFT settings where only a limited number of parameters are trainable. However, finetuning with vanilla PEFT methods overlooks this information, treating all input tokens uniformly and distinguishing their contextual embeddings via an implicit attention mechanism only. 

In this paper, we propose SynMOE, a syntax-aware MOE-guided approach for PEFT-based code intelligence learning, which consists of two stages: (1) Syntactic Information Extraction (SIE) and (2) Syntax-aware MOE-guided Adaptation (SMA). The first stage involves converting source code into ASTs, followed by a systematic extraction of syntactic labels from ASTs, ensuring that each leaf node is associated with a label. However, in current LMs, tokens in code sequences are normally split into diverse sub-tokens according to Byte-Pair Encoding (BPE) \cite{sennrich2016neural} or its variants, making their potential inconsistency with AST leaf nodes and associated syntactic labels. Thus, we further design a token-label alignment algorithm to attach AST-based syntactic labels to each input token of code LMs, thereby facilitating the syntactic information integration in the follow-up stage. In the SMA stage, simply adding the embeddings of both code tokens and syntactic labels is a straightforward way for syntactic information integration, but the latter may overwhelm code tokens' original representations. Therefore, we consider introducing an MOE structure and using syntactic information to affect its combinational logic, realizing the syntax awareness of code token representations. To be specific, SMA consists of a set of lightweight experts, typically trainable low-rank matrices or vectors. Syntactic labels are modeled into high-dimensional embeddings to participate in the routing decision of the MOE's gating network. During the downstream task adaptation, only the parameters of the syntax-aware MOE module are trainable, while the pre-trained parameters of code LMs are frozen, maintaining the parameter efficiency of SynMOE.

We evaluate the effectiveness of SynMOE through a series of experiments on three popular code intelligence tasks: code search, code clone detection, and line-level code completion. Our study spans five code LMs with parameter sizes ranging from millions to billions. We investigate the performance improvements achieved by SynMOE over two representative PEFT methods, LoRA \cite{hu2022lora} and $(IA)^3$ \cite{liu2022few}. In addition, we compare SynMOE with FT as well as several advanced PEFT approaches, such as LoKr \cite{yeh2023navigating}, LoHa \cite{hyeonfedpara}, and AdaLoRA \cite{zhangadaptive}. Experimental results demonstrate that SynMOE consistently outperforms LoRA and $(IA)^3$, with the maximum performance gain of 8.78\%. It also achieves superior performance compared to sophisticated PEFT methods. Furthermore, it achieves results comparable to FT while maintaining the efficiency advantages of less training time and memory consumption. We also design evaluations in the low-resource scenario where vanilla PEFT methods typically perform well, and find that SynMOE remains effective under such conditions, achieving comparable or improved performance with limited training data. It is worth noting that SynMOE is orthogonal to specific PEFT methods, allowing it to be seamlessly integrated with advanced PEFT approaches for further performance gains.

The contributions of this paper can be summarized as: 

\begin{itemize}
\item [1.] We propose a novel SynMOE approach that integrates syntactic code properties via MOE for a more powerful code token modeling, thereby enhancing the performance of code LMs in PEFT settings.
\item [2.] We conduct extensive experiments on three code intelligence tasks with five code LMs. Experimental results demonstrate the effectiveness of SynMOE.
\item [3.] We open-source all the replication packages at https://github.com/ishuoliu/SynMOE to facilitate future research and applications.
\end{itemize}

\section{Preliminary}
Given the broad application and research on LoRA and $(IA)^3$ in both academia and industry \cite{liu2025can, li2024exploring, liu2024moe,zhangadaptive}, we adopt these two PEFT methods in our experiments.

\subsection{LoRA}
\label{LoRA}
LoRA \cite{hu2022lora} is the abbreviation for Low-Rank Adaptation. It hypothesizes that updates to weight matrices have a low intrinsic rank; thus, the updates can be learned efficiently through a smaller subspace. For a pre-trained weight matrix $W_0 \in \mathbb{R}^{d \times k}$, LoRA constrain its update by decomposing the latter into two low-rank matrices that satisfy $W_0 + \Delta W = W_0 + BA$, where $B \in \mathbb{R}^{d \times r}$, $A \in \mathbb{R}^{r \times k}$, and the rank $r \ll min(d, k)$. Initially, $\Delta W = BA$ is zero, which is realized by using a random Gaussian initialization for $A$ and zero for $B$. During tuning, $W_0$ is frozen while only $A$ and $B$ are trainable. Because $r \ll min(d, k)$, the trainable parameters of LoRA are far fewer than those for learning the original $\Delta W$. The principles of LoRA can be applied to any dense layers in the Transformer architecture, but in this paper, we follow the usual practice that applies LoRA to the \textit{query} and \textit{value} projection matrices in the multi-head attention module.

\subsection{$(IA)^3$}

$(IA)^3$ \cite{liu2022few} stands for "Infused Adapter by Inhibiting and Amplifying Inner Activations". Different from LoRA, which focuses on updating the pre-trained weight matrices, $(IA)^3$ explores the element-wise multiplication of models' activations. Specifically, it introduces three learnable vectors, $l_k \in \mathbb{R}^{d_k}$, $l_v \in \mathbb{R}^{d_v}$, and $l_{ff} \in \mathbb{R}^{d_{ff}}$ in each Transformer layer block, rescaling \textit{key} and \textit{value} in the attention mechanism, as well as the intermediate activations of feed-forward networks. Formally, $(IA)^3$ can be represented as:

\begin{equation}
\text{softmax}\left(\frac{Q\left(l_k \odot K^T\right)}{\sqrt{d_k}}\right)\left(l_v \odot V\right)
\end{equation}

\begin{equation}
\left(l_{ff} \odot \gamma\left(W_1 x\right)\right)W_2
\end{equation}

\noindent where $Q$, $K$, and $V$ are the \textit{query}, \textit{key}, and \textit{value} projection matrices, and $d_k$ is the dimension of \textit{key}. $\odot$ denotes the element-wise multiplication. $W_1$ and $W_2$ are weight matrices of the feed-forward network, and $\gamma$ is the feed-forward network nonlinearity. During tuning, in addition to $l_k$, $l_v$, and $l_{ff}$ being trainable, which is an extremely small portion, the others are frozen and do not receive gradient updates.

\section{Methodology}

\begin{figure*}[t]
	\centering
        \setlength{\abovecaptionskip}{0.cm}
		\includegraphics[width=\textwidth]{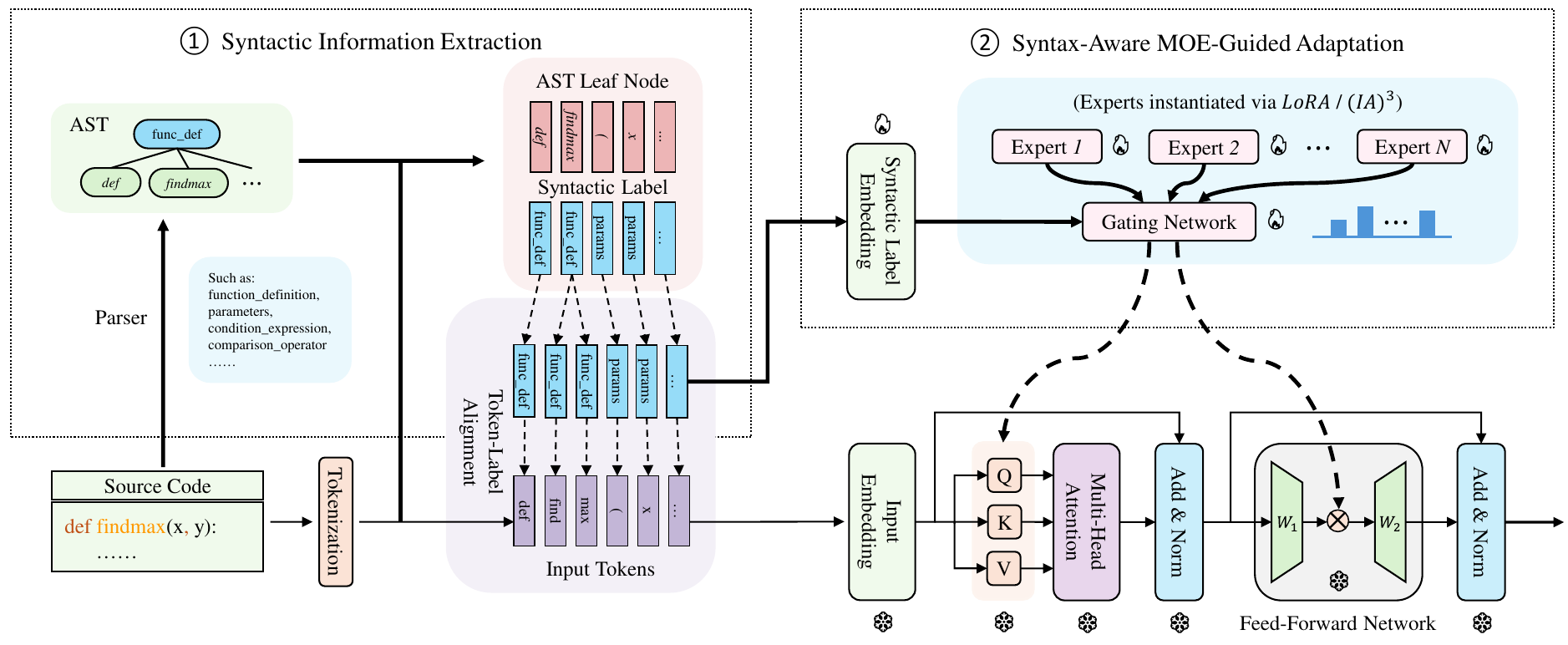}
	  \caption{The workflow of SynMOE. Bold arrows are used to highlight our proposed approach.}\label{framework-figure}
\vspace{-0.4cm}
\end{figure*}

Figure \ref{framework-figure} illustrates the workflow of SynMOE, consisting of two stages: Syntactic Information Extraction (SIE) and Syntax-aware MOE-guided Adaptation (SMA). In the \textbf{first} stage, every token in the source code sequence is assigned a syntactic label according to its position in the associated AST. As for the \textbf{second} stage, each syntactic label is incorporated into a MOE-based PEFT technique (e.g., LoRA and $(IA)^3$) to participate in the encoding of its corresponding code token. As such, enhancing existing PEFT techniques with code modeling capability. The whole tuning process freezes code LMs' pre-trained weights, and only the parameters of the SynMOE portion are updated. In this section, we provide a comprehensive description of the two-stage workflow.

\subsection{Syntactic Information Extraction (SIE)}

This stage involves two sequential procedures: (1) Syntactic label assignment and (2) token-label alignment.

\textbf{Syntactic Label Assignment:} This procedure first transforms a code $S$ to its corresponding AST $T = \left(N, N_{leaf}, r, P \right)$. $N$ represents the collection of non-leaf nodes, while $N_{leaf}$ contains all leaf nodes. Curating leaf nodes according to pre-order traversing can obtain a complete sequence of source code that they are parsed from. $r \in N$ is the root Node, and $P$ indicates the parent-child relationships within the AST. In this paper, we employ tree-sitter \cite{treesitter} to convert code snippets to ASTs. As shown in Figure \ref{ast-figure}, each leaf node is associated with a node type, either \texttt{identifier}, which corresponds to user-defined entities, such as variables (\texttt{x} and \texttt{y}), or the node itself, representing pre-defined entities, such as keywords and operators (\texttt{if} and \texttt{>=}). For the former, we assign their syntactic labels as the types of their parent nodes, such as \texttt{function\_definition} for \texttt{max} while \texttt{parameters} for \texttt{x} and \texttt{y}. For the latter, we group them under a unified label, namely \texttt{<NA>}, since their node values already reflect syntactic roles and there is no need to add extra labels for syntactic identification. As a result of this procedure, each leaf node in the AST is assigned a syntactic label. 

\textbf{Token-Label Alignment:} Although syntactic labels have been obtained and associated with their leaf nodes in ASTs as aforementioned, they cannot be directly mapped with the input tokens of code LMs, which can be summarized into three main sources: first, special tokens (e.g., beginning-of-sequence and end-of-sequence) inserted by LMs' tokenizers are absent in the ASTs; second, the tokenization rules of code tokens are different between code LMs and ASTs; and third, comments that are aggregated into a single node by AST parsers but split into multiple distinct sub-tokens by code LMs. 
The objective of this procedure is to align the extracted syntactic labels with the tokenization results of code LMs, ensuring that each input token has a corresponding syntactic label. Formally, given a source code dataset $S = \{s_n\}_{n=1}^{N_s}$, each code snippet $s_n$ is associated with a joint attribute comprising AST leaf nodes and corresponding syntactic labels $\{(leaf_j, label_j)\}_{j=1}^{N_{leaf}}$. To begin with, we leverage the code LM's tokenizer to convert each source code $s_n$ into input tokens $\{t_i\}_{i=1}^{N_t}$, with $N_t$ denoting the number of input tokens, while initializing two pointers to the initial token and leaf node, respectively, for a left-to-right alignment (Lines 5-6). For each input token $t_i$ to be matched, we first determine whether it is a special token introduced by the tokenizer. If so, a designated label \texttt{<SP>} is assigned (Lines 8-9). Notably, the special tokens vary depending on the specific tokenizer utilized. For instance, the Beginning-Of-Sequence (BOS) and End-Of-Sequence (EOS) tokens are configured as \texttt{<s>} and \texttt{</s>} for CodeBERT, whereas they correspond to \texttt{<|begin\_of\_sentence|>} and \texttt{<|end\_of\_sentence|>} for Deepseek-Coder. Alternatively, if the input token $t_i$ is excluded from the aforementioned categories but falls within a previously identified comment region (Lines 1-4), it is directly labeled as \texttt{<CM>} (Lines 10-11). If the input token $t_i$ is neither a special token nor a token that belongs to a comment, it will be enforced an exact match with the AST leaf node $leaf_j$, so that to assign $leaf_j$'s label $label_j$ to $t_i$. If the exact matching succeed, the increment of the leaf node pointer $j$ will be triggered (Lines 12-14). Otherwise, the $i$th input token will be deemed a subtoken of the $j$th leaf node. In this scenario, a temporary variable $mid$ is maintained to accumulate incoming tokens until the aggregated content precisely aligns with the target leaf node. Once matched, the syntactic label of the leaf node is propagated to all tokens currently held in $mid$, while the leaf node pointer is simultaneously incremented (Lines 15-25). Finally, since input token pointer has been self-incremented during the above loop, it will simply increment one to point to the next input token (Line 26). A detailed overview of this alignment procedure is formalized in Algorithm 1, which ensures that each input token is ultimately assigned an appropriate syntactic label.



\renewcommand{\algorithmicrequire}{\textbf{Input:}}
\renewcommand{\algorithmicensure}{\textbf{Output:}}
\begin{algorithm}[t]
    \caption{Token-Label Alignment}
    \begin{algorithmic}[1]
        \REQUIRE{Source code dataset $\{s_n\}_{n=1}^{N_s}$, Code LM $\Theta$, Pairs of leaf nodes and syntactic labels $\{(leaf_{j}, label_{j})\}_{j=1}^{N_{leaf}}$.}
        \FOR {$s_n$ in $\{s_n\}_{n=1}^{N_s}$}
            \STATE Identify comment regions.
            \STATE Record their start and end positions $\{(p_{start}^c, p_{end}^c)\}_{c=1}^{N_c}$, where $N_c$ denotes the number of comments.
        \ENDFOR
        \STATE Tokenize each source code $s_n$ into input tokens $\{t_i\}_{i=1}^{N_t}$ using the tokenizer of code LM $\Theta$.
        \STATE Initialize input token pointer $i = 1$, and AST leaf node pointer $j=1$.
        \WHILE {$i \leq N_t$}
        \IF {$t_i$ is a special token introduced by the tokenizer}
            \STATE Assign $t_i$ the \texttt{<TK>} label.
        \ELSIF{$t_i$ falls within the comments regions}
            \STATE Assign $t_i$ the \texttt{<CM>} label.
        \ELSIF{$t_i$ exactly matches the $j$-th AST leaf node}
            \STATE Assign $t_i$ the syntactic label of the leaf node.
            \STATE Update pointer $j \leftarrow j + 1$.
        \ELSE 
            \STATE Utilize a variable $mid$ to store the currently unmatched tokens.
            \WHILE {$++i \leq N_t$}
            \STATE $mid \leftarrow mid + t_i$.
            \IF{$mid$ exactly matches the $j$-th AST leaf node}
                \STATE Assign all tokens stored in $mid$ the syntactic label of the $j$-th AST leaf node.
                \STATE Update pointer $j \leftarrow j + 1$.
                \STATE \textbf{break}
                \ENDIF
            \ENDWHILE
        \ENDIF
        \STATE Update pointer $i \leftarrow i + 1$ to process the next input token.
        \ENDWHILE
        \RETURN Pairs of tokens and syntactic labels.
    \end{algorithmic}
\end{algorithm}

\subsection{Syntax-aware MOE-guided Adaptation (SMA)}

Although code tokens carry distinctive syntactic roles, existing PEFT methods treat them uniformly, where their differentiated representations are only derived from an implicit attention mechanism and associated non-linear transformations. Given that the previous stage has attached an explicit syntactic label for each code token, the goal of SMA is to effectively incorporate that syntactic information for more efficient learning with PEFT.

Specifically, the SMA module takes a code sequence embedding and its associated syntactic labels as inputs, denoting as $X \in \mathbb{R}^{L \times D}$ and $S_s \in \mathbb{R}^{L \times N}$, respectively, where $L$ refers to the length of the code sequence, $D$ indicates the model's hidden dimension, and $N$ represents the number of experts of the MOE structure. Afterward, SMA leverages a set of lightweight PEFT techniques (e.g., LoRA or $(IA)^3$) as experts to enhance the modeling capability of a single PEFT module, where each expert is denoted as $\{E_i\}_{i=1}^{N}$. To control the incremental scale of trainable parameters between MOE-based PEFT and its single version, the rank of matrices $A$ and $B$ in LoRA (see Section \ref{LoRA}) is reduced to $\frac{1}{N}$ of the single version. For each $(IA)^3$ expert, it comprises three learnable vectors: $l_k$, $l_v$, and $l_{ff}$. Since the dimension of these vectors is intrinsically dictated by the hidden dimension $d$ of the backbone code LM rather than acting as a flexible hyperparameter, we do not impose additional constraints on the size of individual $(IA)^3$ experts. Subsequently, we combine each expert's representation through the routing weights of a gating network. Recall that each syntactic label was encoded into an $N$-dimension vector; thus, they can also be incorporated into the above routing decision on weight assignment. This procedure can be mathematically expressed as:


\begin{equation}
S_g = softmax(X \cdot W_g) \label{equation-3}
\end{equation}

\begin{equation}
Y = \sum_{i = 1}^{N} (\lambda \cdot S_s + (1 - \lambda) \cdot S_g)\cdot E_i \left(X\right) \label{equation-4}
\end{equation}

\noindent where $W_g \in \mathbb{R}^{D \times N}$ is a trainable weight in the gating network, and $S_g \in \mathbb{R}^{L \times N}$ represents the resulting routing scores. $\lambda$ is a hyperparameter that controls the proportion of $S_s$ and $S_g$, and $S$ refers to the final routing scores that integrate both the syntactic information and the results of the gating network. $E_i \left(X\right)$ represents the forward process in expert $E_i$. The final output $Y$ is computed as a weighted average of the outputs from all experts. 
In this way, SMA enables each token to be encoded by multiple PEFT modules, realizing a more powerful and multi-angle modeling. Meanwhile, associated syntactic information is included to guide the combinatorial modeling process, explicitly introducing the syntax awareness capability of SynMOE. It is worth noting that we do not directly add embeddings of syntactic labels to those of code tokens. Instead, we use a gating network to let the syntactic labels influence the expert routing, because a simple summation would obscure the original representation of code tokens acquired from their code context. When fine-tuning code LMs with SynMOE, we keep code LMs' all parameters frozen while only updating parameters in SynMOE. As for inference, syntactic labels are also required for the pre-filling phase in our experimental setting. 



\section{Experimental Setup}

\subsection{Tasks and Datasets}

To investigate the effectiveness of our proposed SynMOE, we conduct experiments on three typical code tasks from code understanding to generation across diverse PLs, such as Python, Java, and C/C++, thereby ensuring a comprehensive evaluation. Below, we describe the three tasks and associated datasets in detail.

\subsubsection{Code Search (CS)}
In modern software development, code search is one of the most frequent activities \cite{di2023code}. The task aims to retrieve the most relevant code from a large-scale codebase that aligns with the intent of search queries \cite{liu2024empirical}. It measures the semantic relatedness between codes and NL queries within a shared feature space. We conduct experiments on the CosQA dataset \cite{huang2021cosqa}, which contains 20,604 pairs of NL queries and codes, as well as a codebase of capacity 6,267. It collects real user queries from the Microsoft Bing search engine and takes the CodeSearchNet Corpus \cite{husain2019codesearchnet} as a function-level code base. All functions in this dataset are written in Python. It has been officially divided into training, validation, and test sets in the ratio of 19,604:500:500.

\subsubsection{Code Clone Detection (CCD)}
Code clone indicates two code snippets have the same functional semantics but with literal variations \cite{liu2025107775}. It may lead to error propagation through similar codes and severely affect maintenance costs \cite{fang2020functional}. Therefore, code clone detection becomes substantial in software engineering, and it is also fundamental in other tasks like code refactoring and defect prediction. Our experiments use the POJ-104 dataset \cite{mou2016convolutional}, which is collected from a pedagogical programming open judge system, containing 104 problems, with 500 C/C++ programs chosen for each problem. POJ-104 treats all programs solving the same problem as clones, and the objective is to identify whether a given program pair is clone or not. The whole dataset is divided into a training set with 32,000 samples, a validation set with 8,000 samples, and a test set with 12,000 samples.

\subsubsection{Line-level Code Completion (LCC)}
Code completion is regarded as a critical task in promoting efficient software development. It can assist developers in saving time and effort by automatically completing code based on context \cite{guo2023longcoder}. Following previous studies \cite{lu2021codexglue, lu2022reacc}, the task includes two scenarios: token-level code completion and line-level code completion. The former focuses on predicting the next single token, whereas the latter evaluates the quality of the entire generated line. Consequently, line-level code completion is considered more challenging. We use one of the most influential datasets, JavaCorpus \cite{allamanis2013mining}, to perform experiments. JavaCorpus collects over 14 thousand Java projects from GitHub. After filtering by Lu et al. \cite{lu2021codexglue}, the dataset is finally split into 12,934 files for training, 7,189 for validation, and 3,000 for testing.

\subsection{Code LMs and Baselines}

To thoroughly examine our proposed SynMOE, we use five code LMs with diverse parameter scales, which are widely used and well recognized in the community \cite{liu2024delving, zhu2024exploring, ribeiro2022framing}.

$\bullet$ \textbf{CodeBERT} \cite{feng2020codebert} is an encoder-only language model with 125M parameters, pre-trained on a large code corpus. We use it as a base model for code search and code clone detection. 

$\bullet$ \textbf{UniXcoder} \cite{guo2022unixcoder} is a unified cross-modal pre-trained model for programming languages, also containing 125M parameters. Owing to its encoder-decoder structure, it is involved in all three tasks. 

$\bullet$ \textbf{CodeGPT} \cite{lu2021codexglue} has the same model architecture (decoder-only) and training objectives as GPT-2 \cite{radford2019language}. We adopt the 125M-parameter CodeGPT-small-java-adapted for line-level code completion, which uses the GPT-2 as the starting point and is continually trained on 1.6 million Java code. 

$\bullet$ \textbf{Deepseek-Coder} \cite{guo2024deepseek} and \textbf{Qwen2.5-Coder} \cite{hui2024qwen2} are two recently released, well-known large code LMs and have achieved significant advancements. We chose their 1.3B- and 1.5B-parameter models. Owing to their decoder-only structure, they are also applied in the line-level code completion task.

In addition, we also compare SynMOE against the following LoRA-variant methods for a more comprehensive assessment.

$\bullet$ \textbf{LoKr} \cite{yeh2023navigating} is an abbreviation for Low-rank adaptation with Kronecker product. It is an extension of the KronA technique \cite{edalati2022krona} and employs Kronecker products for matrix decomposition. 

$\bullet$ \textbf{LoHa} \cite{hyeonfedpara} stands for Low-rank adaptation with Hadamard product. It finds that decomposing weight updates with the Hadamard product could improve the fine-tuning capability. 

$\bullet$ \textbf{AdaLoRA} \cite{zhangadaptive} assumes that different modules have varying importance. It can adaptively allocate the parameter budget among weight matrices to improve the performance.

\subsection{Evaluation Metrics}

For code search, following previous studies \cite{guo2022unixcoder, ma2023mulcs, shi2023cocosoda}, we adopt the Mean Reciprocal Rank (MRR) as the evaluation metric. MRR computes the average of the reciprocal ranks of the correct code snippets across all queries. A higher MRR score indicates better performance, reflecting that the correct code snippet tends to be ranked more highly. The formula for computing MRR is:

\begin{equation}
\text{MRR} = \frac{1}{|Q|} \sum_{i=1}^{|Q|} \frac{1}{\text{Rank}_i}
\end{equation}

In line with prior works \cite{lu2021codexglue, bui2021self, ding2024traced}, we use the Mean Average Precision (MAP) to evaluate performance for code clone detection. MAP is the mean of average precision scores across all queries. For a single query, its average precision takes into account both the number of relevant objects retrieved and their ranks in the retrieved list.

The evaluation metrics for line-level code completion are Exact Match (EM) and Edit Similarity (ES), which have been widely used in prior works \cite{guo2024ft2ra, liu2024non, zhu2024exploring}. EM evaluates whether the generated output exactly matches the ground truth token-by-token, while ES measures the similarity of the predicted output $\hat{Y}$ and the ground truth $Y$ based on the minimum number of edit operations. It is computed as:

\begin{equation}
\text{ES} = 1 - \frac{Lev(Y, \hat{Y})}{\max(|Y|, |\hat{Y}|)}
\end{equation}

\noindent where $Lev$ denotes the Levenshtein distance, also known as the edit distance.

\subsection{Implementation Details}

The experiments are primarily carried out on an Ubuntu-based GPU server equipped with four RTX 3090 24GB GPUs. The fine-tuning of Deepseek-Coder and Qwen2.5-Coder is instead performed on an RTX A6000 GPU with 48GB memory capacity. All code LMs are obtained from Huggingface\footnote{https://huggingface.co/models}, and we retain their default hyperparameter settings. The LoRA and $(IA)^3$ methods are implemented using the PEFT library\footnote{https://huggingface.co/docs/peft/index}. We perform a grid search over learning rates during fine-tuning. Specifically, we evaluate values in the range of \{5e-5, 2e-5, 1e-5\} for the FT approach, and \{1e-2, 5e-3, 2e-3, 1e-3, 5e-4, 2e-4\} for PEFT approaches. In the code search task, following prior work \cite{guo2022unixcoder}, we set a maximum of 10 training epochs, along with input length limits of 256 tokens for code snippets and 128 tokens for natural language descriptions. For code clone detection, the code LMs are fine-tuned for a maximum of 5 epochs, with an input window size of 400 tokens. In the line-level code completion task, we fine-tune the code LMs for up to 5 epochs, with inputs restricted to a maximum length of 512 tokens. The early stopping strategy with a patience of 3 is applied. Moreover, we set the number of experts uniformly to 4 across experiments and adjust the $\lambda$ value within the interval of $[0, 1]$ to pursue the best results of SynMOE. Further investigation into the impact of the hyperparameters is detailed in Section \ref{section-rq4}. Each experiment is repeated 5 times with different initialization seeds, and the mean and variance are reported.

\begin{table*}[t]
\setlength{\tabcolsep}{4pt}
\renewcommand{\arraystretch}{1.3}
\caption{Evaluation results of SynMOE across three code intelligence tasks. The overall best results (including FT) are indicated in \textbf{bold}, while the best performances among SynMOE and PEFT methods are highlighted with a \colorbox{gray!20}{gray background}. The improvement ratios brought by SynMOE to PEFT methods are indicated in parentheses. Upper-right asterisks denote statistically significant improvements of SynMOE over corresponding PEFT methods. Since all experiments are repeated five times, we report their results in the format of mean${_{\pm std}}$.} \label{RQ1-table}
\centering
\begin{tabular}{ccc|c|cccc}
\hline
\multirow{2}{*}{\textbf{Tasks}} & \multirow{2}{*}{\textbf{Models}} & \multirow{2}{*}{\textbf{Metrics}} & \textbf{FT} & \textbf{LoRA} & \textbf{$(IA)^3$} & \textbf{\makecell[c]{SynMOE + LoRA}} & \textbf{\makecell[c]{SynMOE + $(IA)^3$}} \\ 
& & & 100\% & $\sim$1.84\% & $\sim$0.05\% & $\sim$1.90\% & $\sim$0.41\% \\
\hline
\multirow{2}{*}{\textbf{CS}} & CodeBERT & \multirow{2}{*}{MRR} & \textbf{68.10}${_{\pm0.13}}$ & 64.28${_{\pm0.19}}$$^*$ & 61.49${_{\pm0.19}}$$^*$ & \cellcolor{gray!15}65.78${_{\pm0.17}}$ $(+2.34\%)$ & 65.06${_{\pm0.12}}$ $(+5.80\%)$ \\
& UniXcoder & & \textbf{70.15}${_{\pm0.21}}$ & 67.85${_{\pm0.26}}$$^*$ & 66.17${_{\pm0.16}}$$^*$ & \cellcolor{gray!15}69.14${_{\pm0.25}}$ $(+1.89\%)$ & 67.90${_{\pm0.26}}$ $(+2.62\%)$ \\
\hline
\multirow{2}{*}{\textbf{CCD}} & CodeBERT & \multirow{2}{*}{MAP} & 85.57${_{\pm0.40}}$ & 81.94${_{\pm0.61}}$$^*$ & 80.32${_{\pm0.16}}$$^*$ & \cellcolor{gray!15}\textbf{85.91}${_{\pm0.16}}$ $(+4.85\%)$ & 82.90${_{\pm0.40}}$ $(+3.22\%)$ \\
& UniXcoder & & 90.85${_{\pm0.12}}$ & 90.53${_{\pm0.14}}$$^*$ & 89.91${_{\pm0.23}}$$^*$ & \cellcolor{gray!15}\textbf{92.01}${_{\pm0.39}}$ $(+1.63\%)$ & 91.29${_{\pm0.36}}$ $(+1.53\%)$ \\
\hline
\multirow{8}{*}{\textbf{LCC}} & \multirow{2}{*}{CodeGPT} & EM & \textbf{30.17}${_{\pm0.22}}$ & 28.61${_{\pm0.27}}$$^*$ & 26.21${_{\pm0.47}}$$^*$ & \cellcolor{gray!15}29.97${_{\pm0.18}}$ $(+4.74\%)$ & 28.51${_{\pm0.37}}$ $(+8.78\%)$ \\
& & ES & 62.26${_{\pm0.27}}$ & 61.37${_{\pm0.07}}$$^*$ & 59.00${_{\pm0.49}}$$^*$ & \cellcolor{gray!15}\textbf{62.38}${_{\pm0.19}}$ $(+1.65\%)$ & 61.12${_{\pm0.25}}$ $(+3.59\%)$ \\
& \multirow{2}{*}{UniXcoder} & EM & \textbf{31.87}${_{\pm0.14}}$ & 29.53${_{\pm0.24}}$$^*$ & 27.67${_{\pm0.37}}$$^*$ & \cellcolor{gray!15}30.99${_{\pm0.16}}$ $(+4.93\%)$ & 29.20${_{\pm0.20}}$ $(+5.54\%)$ \\
& & ES & 64.82${_{\pm0.23}}$ & 63.93${_{\pm0.39}}$$^*$ & 61.33${_{\pm0.23}}$$^*$ & \cellcolor{gray!15}\textbf{64.85}${_{\pm0.35}}$ $(+1.43\%)$ & 62.68${_{\pm0.41}}$$(+2.19\%)$ \\
& \multirow{2}{*}{Deepseek-Coder} & EM & \textbf{42.51}${_{\pm0.09}}$ & 41.15${_{\pm0.16}}$$^*$ & 41.41${_{\pm0.12}}$$^*$ & \cellcolor{gray!15}42.34${_{\pm0.22}}$ $(+2.90\%)$ & 42.18${_{\pm0.07}}$ $(+1.85\%)$ \\
& & ES & 73.01${_{\pm0.16}}$ & 72.13${_{\pm0.10}}$$^*$ & 72.28${_{\pm0.16}}$ & \cellcolor{gray!15}\textbf{73.07}${_{\pm0.18}}$ $(+1.29\%)$ & 72.44${_{\pm0.09}}$ $(+0.22\%)$ \\
& \multirow{2}{*}{Qwen2.5-Coder} & EM & 42.87${_{\pm0.19}}$ & 41.30${_{\pm0.29}}$$^*$ & 43.18${_{\pm0.10}}$$^*$ & 43.62${_{\pm0.18}}$ $(+5.62\%)$ & \cellcolor{gray!15}\textbf{43.75}${_{\pm0.14}}$ $(+1.31\%)$ \\
& & ES & 72.80${_{\pm0.12}}$ & 72.68${_{\pm0.20}}$$^*$ & 73.22${_{\pm0.12}}$$^*$ & \cellcolor{gray!15}\textbf{74.05}${_{\pm0.10}}$ $(+1.89\%)$ & 73.96${_{\pm0.09}}$ $(+1.00\%)$ \\
\hline
\end{tabular}
\label{datasets}
\vspace{-0.4cm}
\end{table*}

\section{Experimental Results}

In this section, we conduct extensive experiments to answer the following research questions:

\textbf{RQ1}: What is the effectiveness of SynMOE across three code intelligence tasks?

\textbf{RQ2}: How does SynMOE perform when compared with other advanced PEFT methods?

\textbf{RQ3}: What are the contributions of different components in SynMOE to the performance?

\textbf{RQ4}: What is the impact of different hyperparameters?

\textbf{RQ5}: How effective is SynMOE in the low-resource scenario?

\subsection{RQ1: Overall Effectiveness}

\begin{figure}[t]
	\centering
        \setlength{\abovecaptionskip}{0.cm}
		\includegraphics[width=0.8\textwidth]{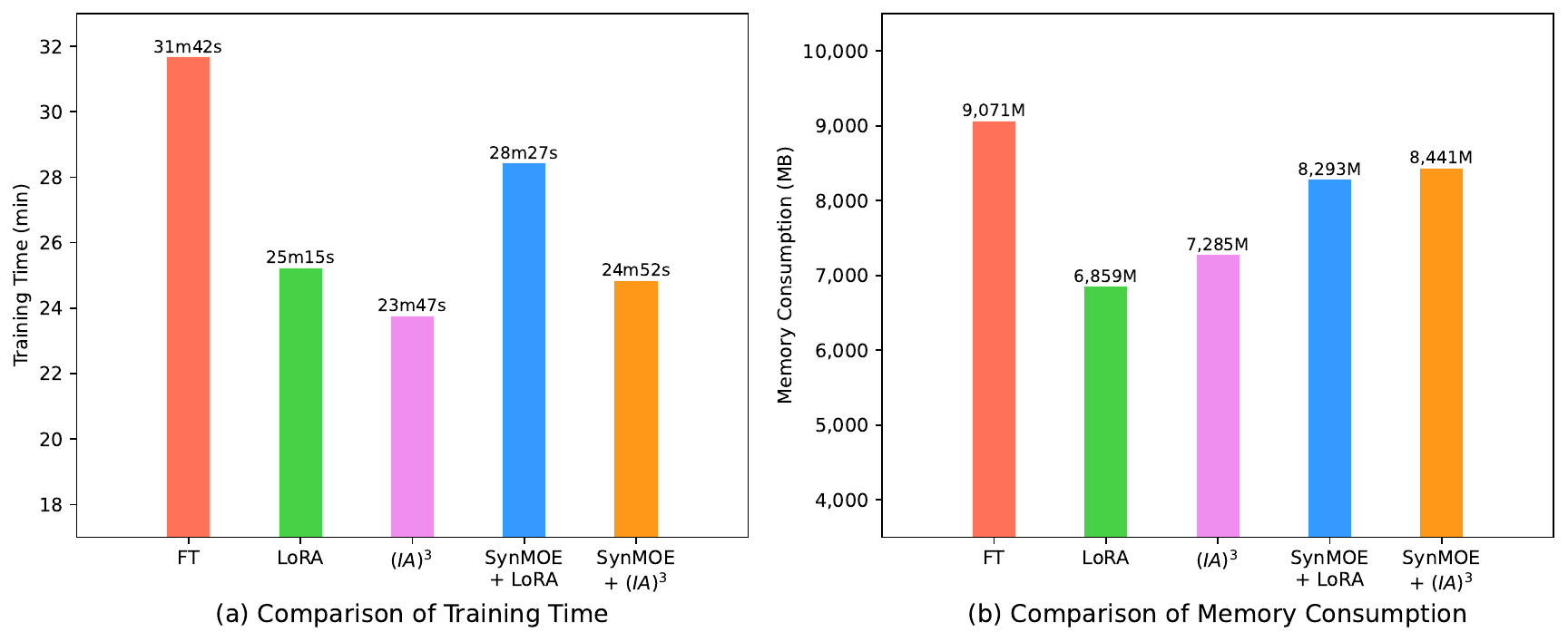}
	  \caption{A comparative analysis of computational overhead for FT, vanilla PEFT methods, and SynMOE.}\label{RQ1-figure}
\vspace{-0.4cm}
\end{figure}

We evaluate the performance of SynMOE in comparison with two representative PEFT methods, LoRA and $(IA)^3$, across three core code intelligence tasks with five representative code LMs. All experiments are repeated five times independently. The full experimental results are summarized in Table \ref{RQ1-table}, where the overall best results (including FT) are indicated in bold, the top-performing values among SynMOE variants and PEFT methods are highlighted with a gray background, and the relative improvement ratios of SynMOE over corresponding vanilla PEFT methods are reported in parentheses. In addition, we analyze the computational efficiency of these methods in Figure \ref{RQ1-figure}. Using CodeGPT as the base model on the line-level code completion task, we measure two key metrics: average training duration per epoch and total memory footprint.

In the code search tasks, SynMOE consistently outperforms PEFT methods across all base models with clear performance gains. For CodeBERT, SynMOE achieves a 2.34\% improvement over LoRA and a substantial 5.80\% gain over $(IA)^3$. For UniXcoder, the gains are 1.89\% over LoRA and 2.62\% over $(IA)^3$. We observe that SynMOE attains the optimal performance when paired with LoRA. This result is intuitive and consistent with LoRA's superior adaptability, which stems from its greater number of trainable parameters. Additionally, SynMOE yields higher gains on CodeBERT compared to UniXcoder. This disparity can be attributed to the different pre-training paradigms of the base models and the advanced adaptation strategies introduced by SynMOE. CodeBERT, as a pioneer code LM, is primarily pre-trained on Masked Language Modeling (MLM) and Replaced Token Detection (RTD) \cite{feng2020codebert}. In contrast, UniXcoder benefits from its unified pre-training objectives and the explicit incorporation of ASTs \cite{guo2022unixcoder}. Consequently, since UniXcoder has already internalized the syntactic priors derived from ASTs during its pre-training phase, the performance gains yielded by SynMOE are somewhat attenuated due to this functional overlap. Furthermore, to validate the significance of the performance difference, we perform Wilcoxon Signed-Rank Tests (WSRT) \cite{wilcoxon1992individual} with a confidence level of 95\% between SynMOE and corresponding PEFT methods. The results in the Table \ref{RQ1-table} specified with upper-right asterisks substantiate that SynMOE statistically significantly outperforms their baselines, e.g., LoRA and $(IA)^3$. 

For code clone detection, SynMOE maintains its superiority across all base models. When evaluated on CodeBERT, SynMOE achieves a 4.85\% improvement over LoRA and a 3.22\% improvement over $(IA)^3$. For UniXcoder, SynMOE yields smaller yet notable gains of 1.63\% and 1.53\% over LoRA and $(IA)^3$, respectively. Aside from the results consistent with those in the preceding code search task, SynMOE achieves even better performance than FT on the code clone detection task. However, FT involves adapting all parameters, while SynMOE only adapts a small proportion, approximately 1.90\% with LoRA and 0.41\% with $(IA)^3$, showing the powerful adaptation efficiency and efficacy of SynMOE.

Regarding line-level code completion, evaluated via Exact Match (EM) and Edit Similarity (ES) across CodeGPT, UniXcoder, Deepseek-Coder, and Qwen2.5-Coder, SynMOE exhibits consistent and considerable gains over PEFT baselines. Across the four base models, SynMOE achieves a maximum improvement of 5.62\% on the EM metric and 1.89\% on the ES metric relative to LoRA. When integrated with $(IA)^3$, SynMOE delivers maximum gains of 8.78\% on EM and 3.59\% on ES, respectively. The findings from the significance test further corroborate that the enhancement brought by SynMOE is statistically significant. Except in the case where SynMOE combined with $(IA)^3$ fails to show significant superiority over vanilla $(IA)^3$ when Deepseek-Coder is used as the base model, SynMOE exhibits significant advantages in the other 15 metric comparisons. Additionally, in terms of the ES metric, the combination of SynMOE and LoRA even outperforms FT, demonstrating its ability to achieve comparable or superior results with notably fewer parameters.

Beyond performance comparisons, the detailed efficiency analysis in Figure \ref{RQ1-figure} reveals that SynMOE strikes a strategic balance in training costs. Although it incurs higher overhead than vanilla PEFT methods, its computational load is lower than that of FT. Given that SynMoE integrates a set of expert modules for syntactic encoding, a marginal increase in training overhead is foreseeable and can be tolerable. Notably, when compared with FT, SynMoE reduces training time by approximately 3-7 minutes per epoch and lowers memory footprint by 600-800 MB. These findings underscore SynMoE's parameter efficiency, demonstrating its ability to deliver robust performance with minimal trainable parameters, thereby making it a highly viable solution for resource-constrained, real-world applications.

\begin{center}
\fcolorbox{black}{gray!10} 
{
\parbox{1\linewidth}
{\textbf{Answer to RQ1}: The proposed SynMOE demonstrates consistent performance improvements over LoRA and $(IA)^3$ across all three code intelligence tasks, with quantified gains ranging from 0.22\% to 8.78\% over PEFT baselines. Compared to FT, SynMOE achieves comparable or superior performance while drastically reducing trainable parameters and computational overhead, establishing it as a practical and efficient approach.}
}
\end{center}

\subsection{RQ2: Comparison with Advanced PEFT Methods}

Our proposed SynMOE has consistently demonstrated effectiveness in enhancing vanilla PEFT methods. In this section, we further validate its capability by comparing it with three advanced PEFT baselines: LoKr, LoHa, and AdaLoRA. These methods are specifically designed to address limitations of vanilla LoRA and improve adaptation performance. We evaluate SynMOE integrated with LoRA against these advanced PEFT methods across three code intelligence tasks, and the full experimental results are summarized in Table \ref{RQ2-table}. The optimal performance is highlighted in bold, and statistically significant improvements of SynMOE over these baselines are marked with asterisks.

As shown in Table \ref{RQ2-table}, SynMOE outperforms or matches these advanced PEFT baselines. In code search and code clone detection, SynMOE achieves an average improvement of 3.73\% over LoKr, with statistical significance validating the superiority of this gain. In contrast, SynMOE exhibits comparable performance against both LoHa and AdaLoRA. Regarding line-level code completion, SynMOE consistently outperforms LoKr, LoHa, and AdaLoRA across all base models and evaluation metrics. Specifically, SynMOE yields marked increases of 12.14\% and 5.11\% over LoKr in terms of EM and ES, respectively. Compared to LoHa, it achieves an average gain of 6.29\% for EM and 3.17\% for ES. Furthermore, it surpasses AdaLoRA by 4.19\% in EM while maintaining a 1.63\% margin in ES, further consolidating its superiority. It is noteworthy that while SynMOE performs on par with LoHa and AdaLoRA on the code search and code clone detection tasks, it delivers a statistically significant performance leap on the line-level code completion task. We hypothesize that this discrepancy stems from the nature of the tasks. For classification tasks, they present a feature space that methods like LoHa and AdaLoRA can already adequately capture, making further performance improvement challenging. In contrast, the generation task involves a more complex learning process within the feature space. By incorporating syntactic information, SynMoE effectively demonstrates its advantage in boosting performance while maintaining strict parameter efficiency. In addition, SynMOE is orthogonal to specific PEFT approaches, thereby enabling seamless integration with these advanced PEFT methods to pursue further performance improvements.

\begin{table*}[t]
\setlength{\tabcolsep}{5pt}
\renewcommand{\arraystretch}{1.3}
\caption{Evaluation results of advanced PEFT methods. The optimal performance is highlighted in \textbf{bold}. Asterisks in the upper right corner denote statistically significant improvements of SynMOE integrated with LoRA over the advanced PEFT methods. Since all experiments are repeated five times, we report their results in the format of mean${_{\pm std}}$.} \label{RQ2-table}
\centering
\begin{tabular}{ccc|cccc}
\hline
\textbf{Tasks} & \textbf{Models} & \textbf{Metrics} & \textbf{LoKr} & \textbf{LoHa} & \textbf{AdaLoRA} & \textbf{SynMOE + LoRA} \\ 
\hline
\multirow{2}{*}{\textbf{CS}} & CodeBERT & \multirow{2}{*}{MRR} & 62.06${_{\pm0.20}}$$^*$ & 65.75${_{\pm0.10}}$ & \textbf{65.95}${_{\pm0.22}}$ & 65.78${_{\pm0.17}}$ \\ 
& UniXcoder & & 67.07${_{\pm0.32}}$$^*$ & \textbf{69.77}${_{\pm0.32}}$ & 68.88${_{\pm0.21}}$ & 69.14${_{\pm0.25}}$ \\
\hline
\multirow{2}{*}{\textbf{CCD}} & CodeBERT & \multirow{2}{*}{MAP} & 82.40${_{\pm0.17}}$$^*$ & 85.82${_{\pm0.20}}$ & 85.79${_{\pm0.31}}$ & \textbf{85.91}${_{\pm0.16}}$ \\
& UniXcoder & & 90.57${_{\pm0.22}}$$^*$ & 91.38${_{\pm0.12}}$ & 91.24${_{\pm0.29}}$ & \textbf{92.01}${_{\pm0.39}}$ \\
\hline
\multirow{8}{*}{\textbf{LCC}} & \multirow{2}{*}{CodeGPT} & EM & 22.83${_{\pm0.32}}$$^*$ & 25.37${_{\pm0.10}}$$^*$ & 27.01${_{\pm0.29}}$$^*$ & \textbf{29.97}${_{\pm0.18}}$ \\
& & ES & 54.87${_{\pm0.18}}$$^*$ & 57.46${_{\pm0.18}}$$^*$ & 60.09${_{\pm0.37}}$$^*$ & \textbf{62.38}${_{\pm0.19}}$ \\
& \multirow{2}{*}{UniXcoder} & EM & 27.73${_{\pm0.09}}$$^*$ & 29.91${_{\pm0.19}}$$^*$ & 30.04${_{\pm0.15}}$$^*$ & \textbf{30.99}${_{\pm0.16}}$ \\
& & ES & 62.24${_{\pm0.34}}$$^*$ & 63.75${_{\pm0.29}}$$^*$ & 64.15${_{\pm0.23}}$ & \textbf{64.85}${_{\pm0.35}}$ \\
& \multirow{2}{*}{Deepseek-Coder} & EM & 41.10${_{\pm0.24}}$$^*$ & 41.57${_{\pm0.34}}$$^*$ & 41.67${_{\pm0.12}}$$^*$ & \textbf{42.34}${_{\pm0.22}}$ \\
& & ES & 71.90${_{\pm0.11}}$$^*$ & 72.07${_{\pm0.44}}$$^*$ & 72.42${_{\pm0.16}}$$^*$ & \textbf{73.07}${_{\pm0.18}}$ \\
& \multirow{2}{*}{Qwen2.5-Coder} & EM & 42.54${_{\pm0.21}}$$^*$ & 42.94${_{\pm0.14}}$$^*$ & 43.17${_{\pm0.17}}$$^*$ & \textbf{43.62}${_{\pm0.18}}$ \\
& & ES & 73.38${_{\pm0.06}}$$^*$ & 73.29${_{\pm0.28}}$$^*$ & 73.53${_{\pm0.07}}$$^*$ & \textbf{74.05}${_{\pm0.10}}$ \\
\hline
\end{tabular}
\label{datasets}
\vspace{-0.4cm}
\end{table*}

\begin{center}
\fcolorbox{black}{gray!10} 
{
\parbox{1\linewidth}
{\textbf{Answer to RQ2}: Our results demonstrate that SynMOE achieves competitive or superior performance compared to state-of-the-art advanced PEFT baselines across all evaluated code intelligence tasks. Critically, because it is orthogonal to specific PEFT methods, SynMOE can be seamlessly integrated with these advanced methods to further boost its effectiveness.}
}
\end{center}

\subsection{RQ3: Ablation Study} 
\label{section-rq3}

We conduct ablation studies to quantify the individual contributions of the two core components in SynMOE, i.e., Syntactic Information Extraction (SIE) and Syntax-aware MOE-guided Adaptation (SMA). The experiments adopt a consistent setup across all base models and code intelligence tasks. We evaluate two ablation configurations: (i) w/o SIE, which involves disabling syntactic information extraction by setting the $\lambda$ value to 0 and relying solely on the gating network for routing; (ii) w/o SMA, which simplifies the MOE module into a single expert setup. Both configurations are implemented using SynMOE integrated with LoRA and $(IA)^3$, and the full results are tabulated in Table \ref{RQ3-table}, where the top-performing results are highlighted in bold.

As shown in Table \ref{RQ3-table}, removing either component consistently results in significant performance degradation, demonstrating the essential and synergistic contributions of each component to overall effectiveness. Specifically, for code search and code clone detection, integrating SynMoE with LoRA shows that omitting SIE and SMA reduces average performance by 1.85\% and 2.21\%, respectively. A similar trend is observed with $(IA)^3$, where the removal of SIE and SMA incurs a performance drop of 1.37\% and 1.46\%, respectively. In the line-level code completion task, removing SIE and SMA incurs average losses of 1.12\% and 1.85\% for LoRA-based SynMOE. The impact persists in $(IA)^3$-based SynMOE, where disabling SIE and SMA results in decrements of 0.91\% and 2.61\%, respectively. It is noteworthy that disabling SMA consistently results in more severe performance degradation than disabling SIE, highlighting SMA's critical role in SynMOE. This can be attributed to the fact that the number of experts is not only critical to the capacity of the MOE module itself, but also influences the effective utilization of syntactic labels. If SynMOE degenerates to a single expert, the syntactic label-based expert routing strategy loses its efficacy, leading to all tokens being treated uniformly under the sole expert. The ablation study suggests that both the syntactic label and the MoE module contribute positively to parameter-efficient fine-tuning.

\begin{table*}[t]
\setlength{\tabcolsep}{5pt}
\renewcommand{\arraystretch}{1.3}
\caption{The ablation study results of SynMOE. The best results in each set of comparisons are highlighted in \textbf{bold}. Since all experiments are repeated five times, we report their results in the format of mean${_{\pm std}}$.} \label{RQ3-table}
\centering
\begin{tabular}{ccc|ccc|ccc}
\hline
\textbf{Tasks} & \textbf{Models} & \textbf{Metrics} & \textbf{\makecell[c]{SynMOE\\+ LoRA}} & w/o SIE & w/o SMA & \textbf{\makecell[c]{SynMOE\\+ $(IA)^3$}} & w/o SIE & w/o SMA \\ 
\hline
\multirow{2}{*}{\textbf{CS}} & CodeBERT & \multirow{2}{*}{MRR} & \textbf{65.78${_{\pm0.17}}$} & 64.92${_{\pm0.10}}$ & 64.98${_{\pm0.31}}$ & \textbf{65.06${_{\pm0.12}}$} & 64.11${_{\pm0.30}}$ & 64.16${_{\pm0.22}}$ \\
& UniXcoder & & \textbf{69.14${_{\pm0.25}}$} & 68.23${_{\pm0.19}}$ & 67.12${_{\pm0.20}}$ & \textbf{67.90${_{\pm0.26}}$} & 67.49${_{\pm0.53}}$ & 66.73${_{\pm0.32}}$ \\
\hline
\multirow{2}{*}{\textbf{CCD}} & CodeBERT & \multirow{2}{*}{MAP} & \textbf{85.91${_{\pm0.16}}$} & 83.50${_{\pm0.63}}$ & 83.77${_{\pm0.69}}$ & \textbf{82.90${_{\pm0.40}}$} & 80.65${_{\pm0.13}}$ & 81.15${_{\pm0.33}}$ \\
& UniXcoder & & \textbf{92.01${_{\pm0.39}}$} & 90.20${_{\pm0.36}}$ & 89.97${_{\pm0.18}}$ & \textbf{91.29${_{\pm0.36}}$} & 90.67${_{\pm0.24}}$ & 90.73${_{\pm0.09}}$ \\
\hline
\multirow{8}{*}{\textbf{LCC}} & \multirow{2}{*}{CodeGPT} & EM & \textbf{29.97${_{\pm0.18}}$} & 29.42${_{\pm0.36}}$ & 28.67${_{\pm0.20}}$ & \textbf{28.51${_{\pm0.37}}$} & 27.66${_{\pm0.08}}$ & 26.30${_{\pm0.32}}$ \\
& & ES & \textbf{62.38${_{\pm0.19}}$} & 61.97${_{\pm0.23}}$ & 61.74${_{\pm0.36}}$ & \textbf{61.12${_{\pm0.25}}$} & 60.49${_{\pm0.32}}$ & 58.91${_{\pm0.17}}$ \\
& \multirow{2}{*}{UniXcoder} & EM & \textbf{30.99${_{\pm0.16}}$} & 30.01${_{\pm0.43}}$ & 30.05${_{\pm0.30}}$ & \textbf{29.20${_{\pm0.20}}$} & 28.63${_{\pm0.23}}$ & 27.99${_{\pm0.11}}$ \\
& & ES & \textbf{64.85${_{\pm0.35}}$} & 63.98${_{\pm0.23}}$ & 63.99${_{\pm0.55}}$ & \textbf{62.68${_{\pm0.41}}$} & 62.37${_{\pm0.13}}$ & 61.68${_{\pm0.14}}$ \\
& \multirow{2}{*}{Deepseek-Coder} & EM & \textbf{42.34${_{\pm0.22}}$} & 42.09${_{\pm0.19}}$ & 41.98${_{\pm0.09}}$ & \textbf{42.18${_{\pm0.07}}$} & 42.03${_{\pm0.18}}$ & 41.29${_{\pm0.19}}$ \\
& & ES & \textbf{73.07${_{\pm0.18}}$} & 72.79${_{\pm0.07}}$ & 72.56${_{\pm0.10}}$ & \textbf{72.44${_{\pm0.09}}$} & 72.42${_{\pm0.11}}$ & 72.20${_{\pm0.16}}$ \\
& \multirow{2}{*}{Qwen2.5-Coder} & EM & \textbf{43.62${_{\pm0.18}}$} & 43.37${_{\pm0.09}}$ & 42.54${_{\pm0.17}}$ & \textbf{43.75${_{\pm0.14}}$} & 43.65${_{\pm0.16}}$ & 43.24${_{\pm0.12}}$ \\
& & ES & \textbf{74.05${_{\pm0.10}}$} & 73.74${_{\pm0.08}}$ & 73.23${_{\pm0.12}}$ & \textbf{73.96${_{\pm0.09}}$} & 73.79${_{\pm0.10}}$ & 73.78${_{\pm0.14}}$ \\
\hline
\end{tabular}
\label{datasets}
\vspace{-0.4cm}
\end{table*}

\begin{center}
\fcolorbox{black}{gray!10} 
{
\parbox{1\linewidth}
{\textbf{Answer to RQ3}: The ablation study quantifies the distinct and synergistic contributions of SIE and SMA to the overall performance. Removing either component leads to consistent degradation across all tasks, with SMA exerting a more pronounced impact.
}
}
\end{center}

\subsection{RQ4: Impact of Hyperparameters} \label{section-rq4}

\begin{figure}[t]
	\centering
        \setlength{\abovecaptionskip}{0.cm}
		\includegraphics[width=1.0\textwidth]{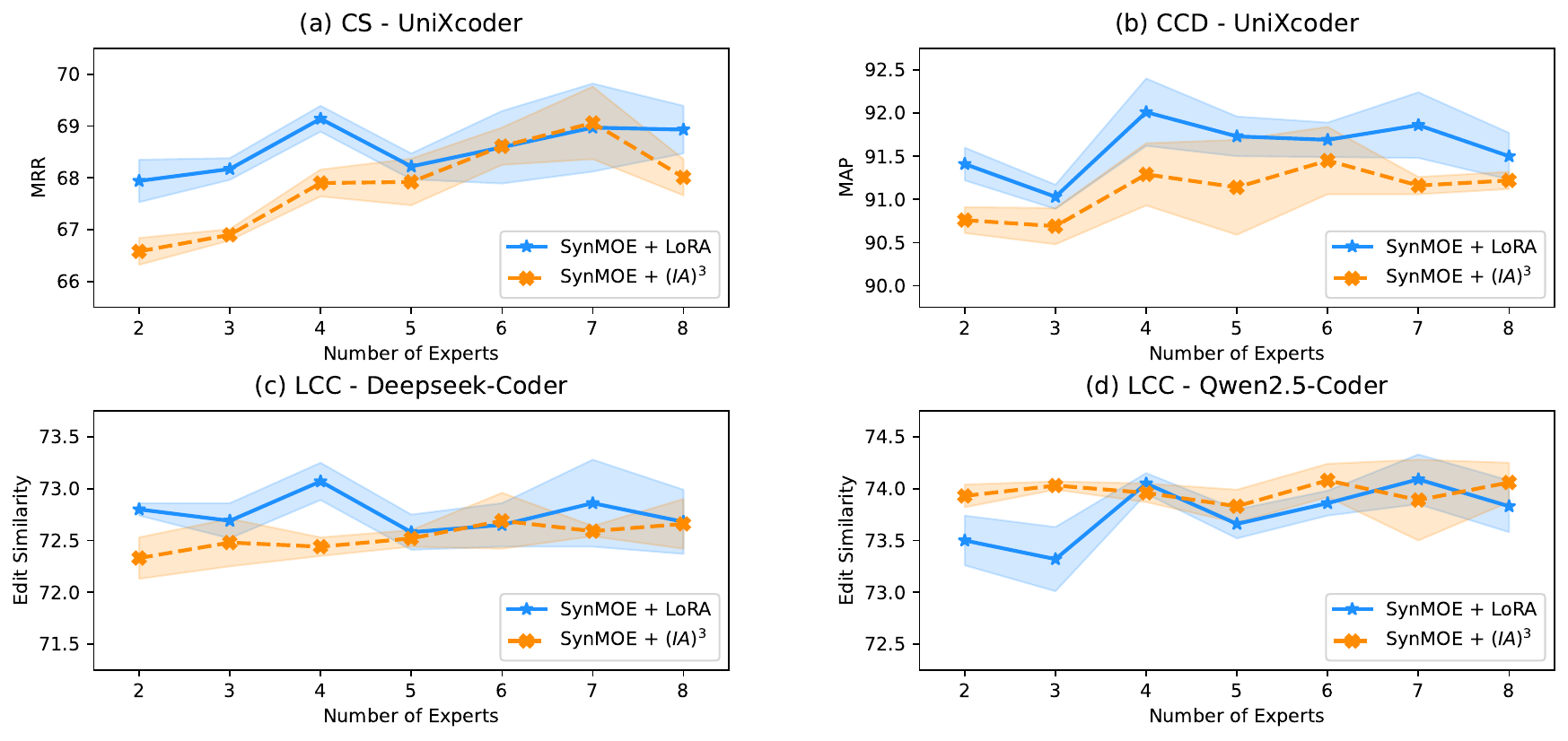}
	  \caption{Parameter analysis of the number of experts. Since all experiments are repeated five times, each data point is specified by an average score, where the score range among repeated experiments is also plotted.}\label{RQ4-figure-1}
\vspace{-0.4cm}
\end{figure}

\begin{figure}[t]
	\centering
        \setlength{\abovecaptionskip}{0.cm}
		\includegraphics[width=1.0\textwidth]{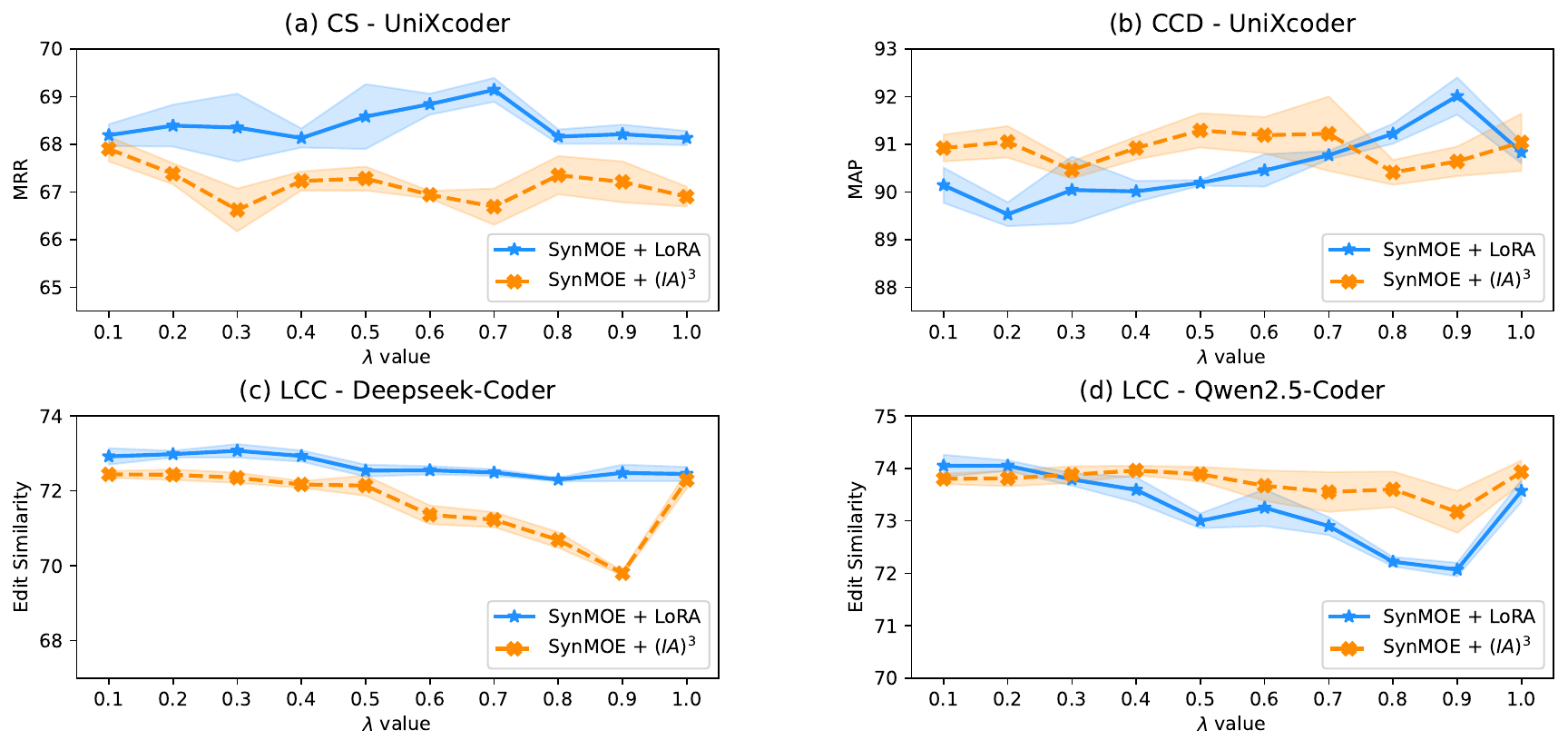}
	  \caption{Parameter analysis of the $\lambda$ value. Since all experiments are repeated five times, each data point is specified by an average score, where the score range among repeated experiments is also plotted.}\label{RQ4-figure-2}
\vspace{-0.4cm}
\end{figure}

This section presents a comprehensive analysis of two critical hyperparameters in SynMOE: the number of experts and the $\lambda$ value in Equation \ref{equation-4}. Intuitively, increasing the number of experts can enhance model capacity by introducing more trainable parameters and greater adaptability. However, an excessive number of experts also incurs higher computational overhead, requiring careful balancing in practice. The $\lambda$ value serves as a control factor to modulate the influence of syntactic labels on the MoE routing process, where a higher $\lambda$ value amplifies the impact of syntactic information on routing decisions. To ensure the generalizability of our findings, we conduct experiments across all three code intelligence tasks, utilizing a diverse suite of base models including UniXcoder, Deepseek-Coder, and Qwen2.5-Coder. Given their outstanding performance in prior experiments, these models serve as logical and representative benchmarks in this study. Both LoRA and $(IA)^3$ integration strategies are evaluated. In each study, we only vary the parameter that needs to be analyzed while keeping the others unchanged.

Regarding the impact of the number of experts, we vary it from 2 to 8 to investigate the scaling effects. We do not set the expert count to one because it's a single expert setup, as previously discussed in Section \ref{section-rq3}. On the other hand, an over-provision of experts incurs prohibitive computational overhead that can even exceed the FT baseline. Since this violates our efficiency prerequisite for SynMOE, we set the maximum number of experts to 8 in our discussion.
As illustrated in Figure \ref{RQ4-figure-1}(a) and \ref{RQ4-figure-1}(b), in the code search and code clone detection tasks, the performance of SynMOE shows an upward trend as the number of experts grows from 2 to 4, consistent across both LoRA and $(IA)^3$ variants. Further increasing the number of experts beyond 4 leads to performance fluctuations, but several improved results are still observed, particularly at 7 experts in the code search task when SynMOE is integrated with $(IA)^3$. For the line-level code completion task (Figure \ref{RQ4-figure-1}(c) and \ref{RQ4-figure-1}(d)), a similar tendency emerges. SynMOE with LoRA achieves peak edit similarity at 4 experts. While SynMOE with $(IA)^3$ exhibits more gradual growth, it also maintains competitive performance at 4 experts. It is also observed that beyond a certain threshold of experts, the performance exhibits fluctuations and is accompanied by increased variance. These results collectively indicate that setting the number of experts to 4 allows SynMOE to strike an effective balance between model performance and computational efficiency. Therefore, we adopt this configuration as the default setting across all experiments. Meanwhile, the performance of SynMOE can be further enhanced by scaling up the number of experts, indicating its underlying potential.

To study the effect of $\lambda$, we sweep its value from 0.1 to 1.0 uniformly in all three code intelligence tasks. As shown in Figure \ref{RQ4-figure-2}, $\lambda$ exhibits a non-monotonic effect on performance. For the code search task (Figure \ref{RQ4-figure-2}(a)), when paired with LoRA, SynMOE achieves its best performance at $\lambda = 0.7$. In contrast, SynMOE with $(IA)^3$ attains its peak performance at $\lambda = 0.1$. In the code clone detection task (Figure \ref{RQ4-figure-2}(b)), the optimal configuration for SynMOE with LoRA is found at $\lambda = 0.9$, while the highest efficacy for SynMOE with $(IA)^3$ is observed when $\lambda$ is set to 0.5. This divergence underscores the specific sensitivity of $\lambda$, where LoRA benefits more from stronger syntactic guidance, while $(IA)^3$ tends to rely more on the contextual information encoding. As for line-level code completion (Figure \ref{RQ4-figure-2}(c) and \ref{RQ4-figure-2}(d)), it is observed that SynMOE maintains a high performance plateau when $\lambda\leq0.4$, but suffers a sharp drop in performance as $\lambda$ approaches 1.0. Specifically, when employing DeepSeek-Coder as the base model, SynMoE with LoRA and $(IA)^3$ reaches its peak efficacy at $\lambda = 0.3$ and $0.1$, respectively. With Qwen2.5-Coder as the base model, the optimal configurations are observed at $\lambda = 0.2$ and $0.4$. Interestingly, certain cases exhibit a rebound in efficacy at $\lambda = 1.0$, suggesting that while syntactic information is inherently beneficial, an excessive routing constraint may partially hinder the model's capabilities. Therefore, finding an optimal equilibrium is paramount for maximizing the overall performance.

\begin{center}
\fcolorbox{black}{gray!10} 
{
\parbox{1\linewidth}
{\textbf{Answer to RQ4}: The performance of SynMOE is affected by both the number of experts and the $\lambda$ value. According to our experiments, setting the number of experts to 4 offers a good trade-off between effectiveness and efficiency. Additionally, task-specific tuning of $\lambda$ is recommended to align with the specific characteristics of target tasks and base models.}
}
\end{center}

\subsection{RQ5: Performance in the Low-Resource Scenario}

\begin{table}[t]
\setlength{\tabcolsep}{3.9pt}
\renewcommand{\arraystretch}{1.3}
\caption{Evaluation results of the code search task and the code clone detection task in the low-resource scenario. The base model used here is UniXcoder. In comparisons of different methods for the same metric, the best results are highlighted in \textbf{bold}. \colorbox{gray!20}{Gray shading} marks the best performance among SynMOE and its PEFT counterparts. Since all experiments are repeated five times, we report their results in the format of mean${_{\pm std}}$.} \label{RQ5-cs}
\centering
\begin{tabular}{ccc|ccccc}
\hline
\textbf{Tasks} & \textbf{Methods} & \textbf{Metrics} & \textbf{1\%} & \textbf{5\%} & \textbf{10\%} & \textbf{25\%} & \textbf{50\%} \\
\hline 
\multirow{4}{*}{\textbf{CS}} & LoRA & \multirow{4}{*}{\textbf{MRR}} & 54.24${_{\pm0.60}}$ & 59.53${_{\pm0.58}}$ & 60.68${_{\pm0.62}}$ & 63.84${_{\pm0.38}}$ & 65.62${_{\pm0.50}}$ \\
& $(IA)^3$ & & 54.55${_{\pm0.34}}$ & 59.66${_{\pm0.41}}$ & 61.33${_{\pm0.30}}$ & 63.85${_{\pm0.42}}$ & 65.27${_{\pm0.74}}$  \\
& SynMOE + LoRA & & \cellcolor{gray!15}\textbf{54.89${_{\pm0.35}}$} & \cellcolor{gray!15}\textbf{61.03${_{\pm0.11}}$} & \cellcolor{gray!15}\textbf{61.63${_{\pm0.39}}$} & \cellcolor{gray!15}\textbf{64.57${_{\pm0.38}}$} & \cellcolor{gray!15}\textbf{66.41${_{\pm0.34}}$}  \\
& SynMOE + $(IA)^3$ & & \cellcolor{gray!15}54.71${_{\pm0.52}}$ & \cellcolor{gray!15}60.13${_{\pm0.29}}$ & \cellcolor{gray!15}61.54${_{\pm0.37}}$ & \cellcolor{gray!15}64.02${_{\pm0.18}}$ & \cellcolor{gray!15}65.93${_{\pm0.41}}$ \\
\hline
\multirow{4}{*}{\textbf{CCD}} & LoRA & \multirow{4}{*}{\textbf{MAP}} & 80.91${_{\pm0.68}}$ & 87.64${_{\pm0.35}}$ & 88.33${_{\pm0.31}}$ & 90.12${_{\pm0.13}}$ & 90.55${_{\pm0.33}}$ \\
& $(IA)^3$ & & 74.33${_{\pm0.19}}$ & 81.64${_{\pm0.24}}$ & 85.44${_{\pm0.26}}$ & 88.13${_{\pm0.18}}$ & 89.49${_{\pm0.18}}$ \\
& SynMOE + LoRA & & \cellcolor{gray!15}\textbf{82.65${_{\pm0.29}}$} & \cellcolor{gray!15}\textbf{88.39${_{\pm0.20}}$} & \cellcolor{gray!15}\textbf{89.78${_{\pm0.33}}$} & \cellcolor{gray!15}\textbf{90.40${_{\pm0.37}}$} & \cellcolor{gray!15}\textbf{91.93${_{\pm0.34}}$}  \\
& SynMOE + $(IA)^3$ & & \cellcolor{gray!15}77.84${_{\pm0.17}}$ & \cellcolor{gray!15}85.98${_{\pm0.49}}$ & \cellcolor{gray!15}88.17${_{\pm0.26}}$ & \cellcolor{gray!15}88.90${_{\pm0.04}}$ & \cellcolor{gray!15}89.93${_{\pm0.15}}$  \\
\hline
\end{tabular}
\label{datasets}
\vspace{-0.4cm}
\end{table}

\begin{table*}[t]
\setlength{\tabcolsep}{6pt}
\renewcommand{\arraystretch}{1.3}
\caption{Evaluation results of the line-level code completion task in the low-resource scenario. The base model used here is Deepseek-Coder. The evaluation metric includes EM and ES. In comparisons of different methods for the same metric, the best results are highlighted in \textbf{bold}. \colorbox{gray!20}{Gray shading} denotes the superior results among SynMOE and PEFT variants within each metric. Since all experiments are repeated five times, we report their results in the format of mean${_{\pm std}}$.} \label{RQ5-lcc}
\centering
\begin{tabular}{ccc|ccccc}
\hline
\textbf{Tasks} & \textbf{Methods} & \textbf{Metrics} & \textbf{1\%} & \textbf{5\%} & \textbf{10\%} & \textbf{25\%} & \textbf{50\%} \\
\hline
\multirow{8}{*}{\textbf{LCC}} & \multirow{2}{*}{LoRA} & \textbf{EM} & \cellcolor{gray!15}\textbf{39.38${_{\pm0.09}}$} & 40.99${_{\pm0.19}}$ & 41.28${_{\pm0.45}}$ & 41.36${_{\pm0.51}}$ & 41.52${_{\pm0.10}}$ \\
& & \textbf{ES} & \cellcolor{gray!15}\textbf{71.23${_{\pm0.10}}$} & 71.99${_{\pm0.13}}$ & 72.04${_{\pm0.12}}$ & 72.29${_{\pm0.42}}$ & 72.31${_{\pm0.09}}$ \\
& \multirow{2}{*}{$(IA)^3$} & \textbf{EM} & 36.92${_{\pm0.02}}$ & 38.60${_{\pm0.15}}$ & 39.61${_{\pm0.12}}$ & 40.61${_{\pm0.11}}$ & 41.26${_{\pm0.06}}$ \\
& & \textbf{ES} & 69.02${_{\pm0.03}}$ & 70.69${_{\pm0.25}}$ & 71.25${_{\pm0.15}}$ & 71.88${_{\pm0.13}}$ & 72.16${_{\pm0.10}}$ \\
& \multirow{2}{*}{SynMOE + LoRA} & \textbf{EM} & 37.06${_{\pm0.03}}$ & \cellcolor{gray!15}\textbf{41.02${_{\pm0.09}}$} & \cellcolor{gray!15}\textbf{41.31${_{\pm0.22}}$} & \cellcolor{gray!15}\textbf{42.02${_{\pm0.22}}$} & \cellcolor{gray!15}\textbf{42.06${_{\pm0.10}}$} \\
& & \textbf{ES} & 69.67${_{\pm0.10}}$ & \cellcolor{gray!15}\textbf{72.24${_{\pm0.30}}$} & \cellcolor{gray!15}\textbf{72.50${_{\pm0.12}}$} & \cellcolor{gray!15}\textbf{72.51${_{\pm0.48}}$} & \cellcolor{gray!15}\textbf{72.84${_{\pm0.31}}$} \\
& \multirow{2}{*}{SynMOE + $(IA)^3$} & \textbf{EM} & \cellcolor{gray!15}37.05${_{\pm0.11}}$ & \cellcolor{gray!15}39.81${_{\pm0.02}}$ & \cellcolor{gray!15}40.43${_{\pm0.24}}$ & \cellcolor{gray!15}41.30${_{\pm0.38}}$ & \cellcolor{gray!15}41.63${_{\pm0.18}}$ \\
& & \textbf{ES} & \cellcolor{gray!15}69.07${_{\pm0.15}}$ & \cellcolor{gray!15}71.31${_{\pm0.17}}$ & \cellcolor{gray!15}71.76${_{\pm0.12}}$ & \cellcolor{gray!15}72.20${_{\pm0.14}}$ & \cellcolor{gray!15}72.36${_{\pm0.09}}$ \\
\hline
\end{tabular}
\label{datasets}
\vspace{-0.4cm}
\end{table*}

Low-resource scenarios are common in real-world software engineering, such as adaptation to new technical stacks or software tasks \cite{liu2023empirical, liu2024delving}, where PEFT methods typically demonstrate superior performance \cite{liu2023empirical, liu2024delving}. In this section, we evaluate SynMOE's capability when training data is scarce across all three code intelligence tasks. Specifically, we utilize UniXcoder for the two classification tasks and Deepseek-Coder for the generation task. This choice is informed by their demonstrated superiority in preceding research questions, which establishes them as representative baseline models. To simulate varying degrees of data scarcity, we randomly sample subsets from the original training partitions, ranging from 1\% to 50\%, while preserving the integrity of the original validation and test sets to ensure unbiased evaluation. The complete results are summarized in Table \ref{RQ5-cs} and Table \ref{RQ5-lcc}.

As shown in Table \ref{RQ5-cs}, SynMOE outperforms vanilla PEFT methods across all data ratios in the two classification tasks. For code search, SynMOE achieves a maximum average boost of 1.65\% in the 5\% setting. While for code clone detection, the improvement peaks at 3.44\% in the 1\% setting. This indicates that the integration of syntactic information not only raises the performance ceiling but also provides an effective inductive bias during data-scarce scenarios, which functions as prior knowledge and provides a shortcut that enables code LMs to swiftly capture the structural patterns of code. Additionally, the standard deviation results reveal that SynMOE maintains a level of stability that is on par with or superior to baseline PEFT approaches. Such a trend suggests that by incorporating syntactic priors, SynMOE stabilizes the learning process and performs more robustly under diverse data-constrained settings.

Regarding line-level code completion, as detailed in Table \ref{RQ5-lcc}, SynMOE yields superior results over vanilla PEFT baselines across nearly all settings, barring a marginal gap for SynMOE+LoRA at the 1\% ratio. The robust performance demonstrates that SynMOE empowers PEFT methods with significant competitiveness in low-resource scenarios. Moreover, SynMOE exhibits superior learning efficiency. With a mere 25\% data allocation, it exceeds the performance of PEFT baselines utilizing double the data volume (50\%). This finding highlights its immense potential for real-world scenarios plagued by the scarcity of large-scale datasets.

\begin{center}
\fcolorbox{black}{gray!10} 
{
\parbox{1\linewidth}
{\textbf{Answer to RQ5}: Across diverse code intelligence tasks, SynMOE exhibits persistent superiority over standard PEFT approaches, effectively outshining them in nearly all low-resource settings. Besides, SynMOE delivers improvements in both learning efficiency and stability. These findings underscore its practical utility in data-scarce real-world scenarios.}
}
\end{center}

\section{Discussion}

\subsection{Case Study}

\begin{figure}[t]
	\centering
        \setlength{\abovecaptionskip}{0.cm}
		\includegraphics[width=0.7\textwidth]{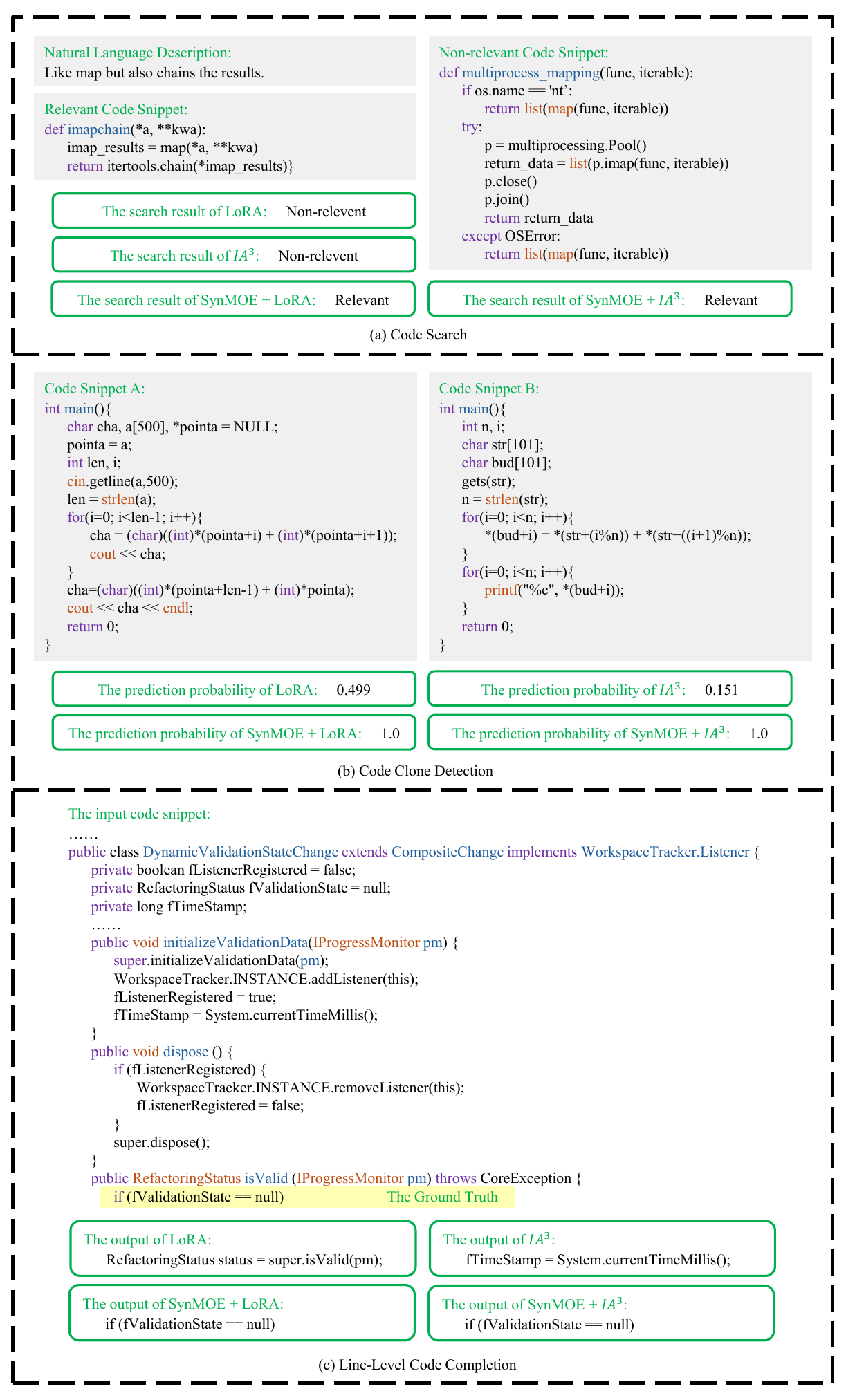}
	  \caption{Case studies in the three code intelligence tasks.}\label{case-study-figure}
\vspace{-0.4cm}
\end{figure}

The advantages of SynMOE come from the syntax-aware MoE module, which incorporates syntactic labels and enables token-wise adaptation. In this section, we conduct case studies to intuitively explore its strength. As shown in Figure \ref{case-study-figure}, we present illustrative examples across three evaluated tasks: code search, code clone detection, and line-level code completion. For each case, we present the results from four approaches, comparing two vanilla PEFT methods against their SynMOE-integrated counterparts.

In our code search case study, given a natural language query describing the target functionality, ``like map but also chains the results'', the objective is to accurately retrieve a relevant code snippet that implement a ``flatMap'' function from a candidate codebase. We report the top-1 retrieval results for different approaches. While LoRA and $(IA)^3$ fail by fetching non-relevant code, the SynMOE-integrated methods successfully identify the correct one. We attribute this to the fact that both LoRA and $(IA)^3$ yield coarse-grained code representations, making them severely misled by the high-frequency appearance of keywords like \texttt{mapping} and \texttt{map} in the non-relevant code snippet, and leading to a spurious high relevance. In contrast, our SynMOE approaches leverage multiple experts to conduct fine-grained analysis, acutely identifying \texttt{itertools.chain} as the pivotal action in \texttt{imapchain}. This establishes a precise alignment with the query phrase ``chains the results'', thereby enabling the successful retrieval of the correct code snippet.

The code clone detection case study takes a pair of code snippets as input, with the objective of determining whether they share identical functional behavior. A preliminary analysis confirms that both code snippets implement identical functionality. Specifically, they both process the input string via "adjacent character addition," where each new character is derived by summing the ASCII codes of the character at the current position and its successor in the original string. The final character is cyclically wrapped around and processed separately after the loop. Although the two code snippets differ significantly in variable names (\texttt{cha}/\texttt{pointa}/\texttt{len} vs. \texttt{bud}/\texttt{str}/\texttt{n}), I/O functions (\texttt{cin}/\texttt{cout} vs. \texttt{gets}/\texttt{printf}), and loop structures, their core logic remains perfectly isomorphic. Nevertheless, vanilla LoRA yields a prediction probability of only 0.499, failing to cross the classification threshold, while $(IA)^3$ struggles even more severely, with a confidence score of just 0.151. These poor scores indicate that vanilla PEFT methods are easily distracted by lexical differences, token mismatches, and structural variations. Conversely, SynMOE-integrated methods achieve a perfect prediction probability of 1.0, demonstrating they focus on underlying dependencies and enabling code LMs to robustly align the core patterns of the two code snippets.


For the line-level code completion instance, it contains an incomplete code snippet and the corresponding ground truth as input, defining a class named \texttt{DynamicValidationStateChange}, which is part of the Eclipse Ruby Development Tools plugin. It serves as a state monitor to manage the dynamic validation state of refactoring operations. It contains two implemented functions, \texttt{initializeValidationData} and \texttt{dispose}. The former aims to register a listener and record the current timestamp, while the latter handles resource disposal and cleanup. The function to be implemented, \texttt{isValid}, is responsible for validity checking. As shown in the Figure \ref{case-study-figure}(c), the ground truth for \texttt{isValid} is simply the conditional check: \texttt{if (fValidationState == null)}. The outputs of both SynMOE variants correctly produce this conditional, aligning well with the ground truth. In contrast, the LoRA output: \texttt{RefactoringStatus status = super.isValid(pm);}, misses the core logic entirely. It does not check \texttt{fValidationState} and instead directly calls the superclass method, which does not fulfill the required behavior. As for $(IA)^3$, it also fails to understand the intent of the code and generates irrelevant elements, such as the time stamp. 

\subsection{Threats to validity}

In this section, we discuss potential threats to the validity of this work from two perspectives.

\textbf{External Validity.} We identify three major threats to the external validity, stemming from the limited selection of evaluation tasks, base models, and PEFT methods. Although we conduct experiments on three representative code intelligence tasks, including the objective of both code comprehension and code generation, other important tasks, such as code translation \cite{10.1145/3728940, xue2025translibeval} and program repair \cite{wu2026debugrepair, li2024exploring}, are not included in our evaluation. Further investigation into the performance of SynMOE on a broader range of code intelligence tasks is still needed. Additionally, the largest code LM employed in our experiments has approximately 1.5 billion parameters. The effectiveness of SynMOE on even larger models is yet to be investigated. Moreover, we evaluate the improvements of SynMOE over two widely-used PEFT methods in this work, namely LoRA and $(IA)^3$. As SynMOE is orthogonal to most PEFT approaches, including diverse adapters \cite{houlsby2019parameter} and LoRA variants \cite{yeh2023navigating, hyeonfedpara, zhangadaptive}, its performance in these configurations is worth further exploration. For the above three potential threats, we will design follow-up experiments to explore them in the future.

\textbf{Internal Validity} primarily concerns factors that could impact the experimental results, such as the hyperparameter settings. In our experiments, hyperparameters are used in both the code LMs and their adaptation on downstream tasks. To reduce potential threats introduced by different model configurations, all code LMs used in this study are obtained from Huggingface with their default configurations. For the hyperparameters related to the adaptation process, we consistenly follow the settings used in prior works, including training epochs and maximum input length. We also perform a grid search over learning rates to obtain the best results. In addition, we conduct a detailed hyperparameter analysis on the number of experts and the $\lambda$ value to uncover their influence on performance. Despite the difficulty in identifying optimal hyperparameter settings, our effort can significantly mitigate these threats.

\section{Related Work}
\subsection{Parameter-Efficient Fine-Tuning in Code Intelligence}

Code Language Models (code LMs) have achieved significant accomplishments in many code intelligence tasks, such as code search \cite{park2023contrastive, xie2023survey}, code clone detection \cite{du2024adaccd, li2023zc}, and code completion \cite{guo2024ft2ra, izadi2024language}. A typical approach to adapting code LMs to downstream tasks is FT. However, it becomes computationally expensive and memory-intensive with the continuous increase in the scale of code LMs. Therefore, some PEFT methods are proposed to alleviate the issues in the field of Artificial Intelligence (AI). Adapter tuning \cite{houlsby2019parameter} adds compact bottleneck modules (adapters) to each layer of the Transformer architecture \cite{vaswani2017attention}, fine-tuning these new modules solely and freezing the other parameters. LoRA \cite{hu2022lora} takes inspiration from the hypothesis that the change in weights during model adaptation has a low intrinsic rank. It learns parameter updates by trainable rank decomposition matrices. Prompt tuning \cite{lester2021power} applies learnable vectors at the initial word embedding layer, aligning the objective of downstream tasks with that in the pre-training stage, and thus achieving decent performance with very few parameters. Similarly, prefix tuning \cite{li2021prefix} introduces learnable vectors to the \textit{key} vector and the \textit{value} vector across all Transformer layers. $(IA)^3$ \cite{liu2022few} introduces three learnable rescaling vectors to rescale the \textit{key} vector, \textit{value} vector, and the activations of feed-forward networks. The rescaling is realized by the element-wise multiplication operation, and due to its simplicity, $(IA)^3$ introduces little computational overhead.

To be specific, in the field of code intelligence, Wang et al. \cite{wang2022no} first empirically evaluate the effect of prompt tuning. Another work \cite{wang2023one} concentrates on adapter tuning and uncovers its advantages in low-resource scenarios. Li et al. \cite{li2024exploring} conduct four PEFT methods on four code LMs that have sizes around 7B or larger, exploring their capabilities on automated program repair. Liu et al. \cite{liu2024delving} investigate adapter tuning and LoRA for tasks related to code changes. They find that both PEFT methods excel in classification tasks, but do not gain an advantage in generative tasks. Focusing on code-change learning, their follow-up work \cite{liu2023empirical} proposed Pasta$_K$, which combines PEFT and FT methods across different layers, achieving comparable or superior performance to FT, while keeping parameter efficiency. However, Pasta$_K$ is only applicable for code-change-related tasks, because its layer-wise combination is guided by their proposed probing tasks for code-change learning. In addition, to exploit the deep semantics embedded in the formal structure of code, Yang et al. \cite{yang2026parameter} proposed Graph-LoRA, a PEFT variant based on the Attributed Patch Semantic Graph (APSG) for exclusively modeling code diffs in patch verification. Yu et al. \cite{yu2025mixture} point out that existing PEFT methods struggle to capture common features shared across languages. Consequently, for the task of multilingual code summarization, they develop a LoRA-based approach incorporating a universal expert and a set of specialized linguistic experts to achieve superior performance. To date, existing works either fail to explicitly incorporate code features or focus solely on a single, specific task. In this paper, we integrate AST-based syntactic labels into vanilla PEFT methods, enhancing their code understanding performance while maintaining parameter efficiency.

\subsection{Mixture-of-Experts}

The concept of mixture-of-experts was first introduced by Jacob et al. \cite{jacobs1991adaptive}, applying multiple sub-modules (experts), each of which is trained with a specific expertise. In the era of LLMs, MOE is incorporated with the Transformer architecture, introducing several sparse experts to the feed-forward sub-layer and using a trainable gating network to assign weights of experts on a per-example basis. It has shown promising effectiveness \cite{du2022glam, mustafa2022multimodal, jiang2024mixtral} but is still associated with high computational costs as FT. Recently, researchers have begun to combine the concept of MOE with PEFT methods. Liu et al. \cite{liu2024moe} explore the potential of using LoRAs as experts for multi-task medical applications. Similarly, Wu et al. \cite{wumixture} delve into the gating network, investigating how to dynamically and efficiently compose multiple trained LoRAs while preserving their individual characteristics. Wang et al. \cite{wang2022adamix} leverage a mixture of adapters to improve the performance of natural language understanding and generation tasks. Some other techniques, such as introducing asymmetric low-rank adaptation \cite{wang2024malora} or assigning more LoRA experts to higher layers \cite{gao2024higher}, have also been explored. In this paper, we propose to leverage MOE to optimize the overall performance of PEFT methods by incorporating syntactic characteristics in the context of code intelligence.

\section{Conclusion}

In this paper, we propose SynMOE to incorporate token-wise syntax awareness into existing PEFT methods through a mixture-of-experts, thereby fostering their learning efficacy in the code intelligence field. In experiments, we examine the performance of SynMOE across three code intelligence tasks using five code LMs, focusing on its improvements over two widely adopted PEFT approaches, LoRA and $(IA)^3$. Experimental results show that SynMOE achieves consistent performance improvements. Moreover, it also outperforms several advanced PEFT methods. As for the comparison with FT, SynMOE achieves comparable or superior performance with reduced training time and lower memory consumption. In our future work, more PEFT methods could be integrated into SynMOE to further enhance the performance of code LMs under PEFT settings. 


\begin{acks}
This work was partially supported by the National Natural Science Foundation of China (Grant Nos. 62502283 and U24B20149), the Natural Science Foundation of Shandong Province (Grant No. ZR2024QF093), and the Young Talent of Lifting Engineering for Science and Technology in Shandong, China (Grant No. SDAST2025QTB031). Besides, it is partially supported by the General Research Fund of the Research Grants Council of Hong Kong and the research funds of the City University of Hong Kong (Grant Nos. 6000796, 9229109, 9229098, 9220103, 9229029).
\end{acks}

\bibliographystyle{ACM-Reference-Format}
\bibliography{sample-base}

\end{document}